\documentclass[aps,prd,preprint,eqsecnum,nofootinbib,groupedaddress,showpacs,floatfix]{revtex4}

\usepackage{graphicx,epsf,color,amsmath}
\newif\ifdraft
\drafttrue
\newif\ifpreprint
\def\fig#1{Fig.~{\ref{#1}}}
\def\Fig#1{Fig.~{\ref{#1}}}

\def\Sect#1{Sect.~{\ref{#1}}}

\def\cseps{\delta}
\def\tlambda{{\tilde \lambda}}
\def\cg{c_\Gamma}
\def\spa#1.#2{\left\langle#1\,#2\right\rangle}
\def\spb#1.#2{\left[#1\,#2\right]}
\def\div{{\rm div.}}
\def\ldiv{{\rm lead.\ div.}}
\def\tree{{\rm tree}}
\def\grav{{}}
\def\glue{{\rm YM}}
\def\pol{\varepsilon}

\def\eps{\epsilon}

\def\eqn#1{Eq.~(\ref{#1})}

\def\eqns#1#2{Eqs.~(\ref{#1}) and~(\ref{#2})}

\def\oneloop{{\rm 1\hbox{-}loop}}
\def\twoloop{{\rm 2\hbox{-}loop}}
\def\Lloop{{L \hbox{-}\rm loop}}
\def\Lmoneloop{{(L-1) \hbox{-}\rm loop}}
\def\Lmtwoloop{{(L-2) \hbox{-}\rm loop}}
\def\tree{{\rm tree}}
\def\oneloop{{\rm 1\hbox{-}loop}}
\def\Ord{{\cal O}}
\newbox\charbox
\newbox\slabox
\def\s#1{{      % Feynman slash
        \setbox\charbox=\hbox{$#1$}
        \setbox\slabox=\hbox{$/$}
        \dimen\charbox=\ht\slabox
        \advance\dimen\charbox by -\dp\slabox
        \advance\dimen\charbox by -\ht\charbox
        \advance\dimen\charbox by \dp\charbox
        \divide\dimen\charbox by 2
        \raise-\dimen\charbox\hbox to \wd\charbox{\hss/\hss}
        \llap{$#1$} }}

\begin{document}

\ifpreprint
UCLA/14/TEP/102 \hfill $\null\hskip 4cm \null$  \hfill
\fi

\title{On Loop Corrections to Subleading Soft Behavior of Gluons \\ and
  Gravitons\\}

\author{Zvi~Bern, Scott~Davies and Josh~Nohle}
\affiliation{
Department of Physics and Astronomy\\
University of California at Los Angeles\\
Los Angeles, CA 90095-1547, USA\\
\vskip 1 cm }

\begin{abstract}
Cachazo and Strominger recently proposed an extension of the
soft-graviton theorem found by Weinberg. In addition, they proved the
validity of their extension at tree level. This was motivated by a
Virasoro symmetry of the gravity $S$-matrix related to BMS symmetry.
As shown long ago by Weinberg, the leading behavior is not corrected
by loops. In contrast, we show that with the standard definition of
soft limits in dimensional regularization, the subleading behavior
is anomalous and modified by loop effects.  We argue that there are no
new types of corrections to the first subleading behavior beyond one
loop and to the second subleading behavior beyond two loops.  To
facilitate our investigation, we introduce a new momentum-conservation
prescription for defining the subleading terms of the soft limit.  We
discuss the loop-level subleading soft behavior of gauge-theory
amplitudes before turning to gravity amplitudes.
\end{abstract}

\pacs{04.65.+e, 11.15.Bt, 11.25.Db, 12.60.Jv \hspace{1cm}}

\maketitle

\section{Introduction}

Recent years have seen enormous advances in our ability to calculate
scattering amplitudes in gauge and gravity theories.  These advances
allow us to address various fundamental issues in such theories.  Some
time ago Weinberg presented a theorem for the universal factorization
of scattering amplitudes when gravitons become soft~\cite{Weinberg}.
Recently Weinberg's soft-graviton theorem was shown to be a Ward
identity~\cite{Strominger} of the Bondi, van der Burg, Metzner and
Sachs (BMS)~\cite{BMS} symmetry.  Along these lines, Strominger
conjectured that an extension of Weinberg's
theorem~\cite{StromingerNotes} for the first subleading terms in the
soft limit follows from BMS symmetry.  Supporting evidence has been
presented recently by Cachazo and Strominger~\cite{CachazoStrominger},
proving that it holds at tree level. Interestingly, Cachazo and
Strominger also showed that the second-order subleading correction to
the tree behavior is also universal.  These results are similar to the
universal subleading soft-photon behavior proven long ago by
Low~\cite{Low}.  The first subleading soft-graviton behavior was first
discussed by White using eikonal methods~\cite{WhiteGrav}.  Very
recently, the subleading soft behavior at tree level has also been
shown to be universal outside of four dimensions~\cite{Volovich}.

One might hope that at least the first subleading soft behavior is a
theorem valid to all loop orders, as suggested by its link to BMS
symmetry~\cite{CachazoStrominger}. However, symmetries at loop level
are delicate because of the need to regularize ultraviolet and
infrared divergences.  The required regularization can modify Ward
identities derived from symmetries.  In this paper, we demonstrate in
a simple way that graviton infrared singularities imply that there are
loop corrections to the subleading behavior of scattering amplitudes
as external gravitons become soft, when we use the standard
definition of such limits.  These corrections are
effectively a quantum breaking of the symmetry responsible for the
tree-level behavior.

In order to understand the loop-level behavior of soft gravitons, it
is useful to first look at the well-studied case of loop corrections
to soft gluons in quantum chromodynamics
(QCD)~\cite{OneLoopSoftBern,OneLoopSoftKosower}.  The subleading
soft-gluon behavior was already discussed using the eikonal
approach~\cite{WhiteYM}.  A simple proof of the universal subleading
soft behavior of gluons at tree level was recently
given~\cite{SoftGluonProof}, following the corresponding proof for
gravitons~\cite{CachazoStrominger}.  The connection between the two
theories is not surprising.  Gravity scattering amplitudes are closely
related to gauge-theory ones and can even be constructed directly from
them~\cite{KLT,BGK,BDDPR,OneLoopMHVGrav,BCJ}.

At one loop, the modifications to the leading soft-gluon behavior are
directly tied to the infrared singularities, and can be used
to deduce the complete correction including finite
parts~\cite{OneLoopSoftBern}.  When a gluon becomes soft, there is a
mismatch between the infrared singularities at $n$ points and at $n-1$
points, so loop corrections to the soft function are required to
absorb this mismatch.  Following
the gauge-theory case, we use the infrared singularities of gravity
loop amplitudes~\cite{Weinberg,GravityIR} to deduce the existence of
loop corrections to the subleading soft-graviton behavior.
As in QCD, discontinuities in the infrared singularities arise as one goes from
$n$ points to $n-1$ points by taking a soft limit in the standard way. 
In gravity, the leading soft-graviton behavior is smooth because the
dimensionful coupling ensures that any discontinuity is suppressed by
at least one additional factor of the soft
momentum~\cite{OneLoopMHVGrav}.  However, since there is less
suppression in subleading soft pieces, loop corrections survive.  This
allows us to demonstrate in a simple way that the subleading behavior
of gravitons indeed has loop corrections similar to the loop corrections
that appear in QCD.  As the loop order increases, the
suppression increases. Hence, the first subleading behavior is
protected against corrections starting at two loops and the second
subleading behavior is protected against corrections starting at three
loops.

This paper is organized as follows.  In \Sect{PreliminariesSection},
we give preliminaries on the tree-level behavior of soft gluons and
gravitons.  In \Sect{LoopSection}, we turn to the main subject of this
paper: the behavior of the subleading contributions at loop level,
showing that there are nontrivial one-loop corrections to subleading
soft-graviton behavior. In \Sect{AllLoopSection}, we discuss the
all-loop behavior.  We give our conclusions in
\Sect{ConclusionSection}.

\section{Preliminaries}
\label{PreliminariesSection}

%%%%%%%%% FIGURE %%%%%%%%%%%%%%%                                              
\begin{figure}[tb]
\begin{center}
\vskip .7 cm 
\includegraphics[scale=.4]{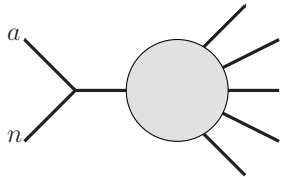}
\end{center}
\vskip -.7 cm 
\caption[a]{\small The diagrams where leading and subleading contributions
  to the tree soft factor arise.  Leg $n$ is the soft leg.
\label{TreeSoftFigure}
}
\end{figure}
%%%%%%%%%%%%%%%%%%%%%%%%%%%%%%%%

In this section, we summarize the soft behavior of gravitons and
gluons at tree level, including their subleading behavior.

\subsection{Soft gravitons}

At tree level, consider the soft scaling of momentum $k_n$ of 
an $n$-point amplitude,
\begin{equation}
k_n^{\alpha\dot \alpha} \rightarrow \cseps k_n^{\alpha\dot \alpha} \,, \hskip 2 cm 
\lambda_n^\alpha \rightarrow \sqrt{\cseps} \lambda_n^\alpha\,, \hskip 2 cm 
\tlambda_n^{\dot \alpha} \rightarrow \sqrt{\cseps} \tlambda_n^{\dot \alpha}\,,
\label{SoftKinematicsReal}
\end{equation}
where $k_n^{\alpha\dot \alpha} = \lambda^\alpha_n \tlambda^{\dot
  \alpha}_n$ is the standard decomposition of a massless momentum in
terms of spinors. (See e.g. Ref.~\cite{LanceTasi} for the
spinor-helicity formalism used for scattering amplitudes.)  In the
limit (\ref{SoftKinematicsReal}), an $n$-point graviton tree amplitude behaves
as~\cite{CachazoStrominger}
\begin{equation}
M_{n}^\tree \rightarrow
\Bigl( \frac{1}{\cseps} \, S_n^{(0)\grav} 
       +  S_n^{(1)\grav}
       + \cseps \, S_n^{(2)\grav} \Bigr) M_{n-1}^\tree + \Ord(\cseps^2) \,,
\label{GeneralGravSoftLimitReal}
\end{equation}
where $\cseps$ is taken to be a small parameter.
The soft operators are
\begin{eqnarray}
&& S_n^{(0)\grav} =  \sum_{i = 1}^{n-1} \frac{\pol_{\mu\nu} k_i^\mu k_i^\nu}
            {k_n \cdot k_i}\,, \nonumber \\
&& S_{n}^{(1)\grav} = \sum_{i=1}^{n-1} \frac{\pol_{\mu\nu} k_i^\mu k_{n\rho} J_i^{\rho\nu}}
                                {k_n \cdot k_i} \,, \nonumber \\
&& S_{n}^{(2)\grav} = \frac{1}{2} 
       \sum_{i=1}^{n-1} \frac{\pol_{\mu\nu} k_{n\rho} J_i^{\rho\mu}
                                       k_{n \sigma} J_i^{\sigma\nu}}
                                {k_n \cdot k_i} \,,
\label{SubleadingSoftGrav}
\end{eqnarray}
where $\pol_{\mu\nu}$ is the graviton polarization tensor of the soft
leg $n$ and $J_i^{\mu\nu}$ is the angular momentum operator for
particle $i$.  $S_n^{(0)\grav}$ is the leading term found long ago by
Weinberg~\cite{Weinberg}.  
For simplicity, we suppress powers of the gravitational coupling
$\kappa/2$ here and in the remaining part of the paper. 
In a helicity basis with a plus-helicity
soft graviton, the explicit forms of the operators are
\begin{eqnarray}
 S_n^{(0)\grav} & =&  -\sum_{i = 1}^{n-1} 
 \frac{\spb{n}.i \spa{x}.i \spa{y}.i}{\spa{n}.{i} \spa{x}.{n} \spa{y}.{n}}\,,
   \nonumber \\
S_{n}^{(1)\grav} &= &-\frac{1}{2} \sum_{i=1}^{n-1} \frac{\spb{n}.{i}} {\spa{n}.{i}}
   \biggl(\frac{\spa{x}.i}{\spa{x}.n} + \frac{\spa{y}.i}{\spa{y}.n} \biggr)
   \tlambda^{\dot\alpha}_n \frac{\partial}{\partial\tlambda^{\dot\alpha}_i} \,,
      \nonumber\\
S_{n}^{(2)\grav} &=& -\frac{1}{2} \sum_{i=1}^{n-1} \frac{\spb{n}.{i}} {\spa{n}.{i}}
           \tlambda^{\dot\alpha}_n\tlambda^{\dot\beta}_n
     \frac{\partial^2}{\partial \tlambda_i^{\dot\alpha} 
                       \partial \tlambda_i^{\dot\beta} } \,,
\end{eqnarray}
where $\lambda_x$ and $\lambda_y$ are arbitrary massless reference
spinors, which reflect gauge invariance.  We follow the standard
conventions of $s_{ab} = \spa{a}.b \spb{b}.a$.  The case of a
minus-helicity soft graviton follows from parity conjugation.
The first subleading behavior was discussed first in Ref.~\cite{WhiteGrav}.

It is convenient to present the subleading behavior in terms of a
holomorphic scaling of the spinors~\cite{CachazoStrominger}.  An
advantage is that it makes the factorization channels
clearer because the universal subleading behavior appears as poles in
the scattering amplitudes.  Taking leg $n$ of an $n$-point amplitude
to be a soft plus-helicity graviton, we scale the spinors as
\begin{equation}
k_n^\mu \rightarrow \cseps k_n^\mu\,, \hskip 2 cm 
\lambda_n^\alpha \rightarrow \cseps \lambda_n^\alpha\,, \hskip 2 cm 
\tlambda_n^{\dot \alpha} \rightarrow \tlambda_n^{\dot \alpha}\,.
\label{SoftKinematics}
\end{equation}
Under this rescaling,
tree-level graviton amplitudes behave as~\cite{CachazoStrominger}
\begin{equation}
M_{n}^\tree \rightarrow
\Bigl(   \frac{1}{\cseps^3} S_{n}^{(0)\grav} 
       + \frac{1}{\cseps^2} S_{n}^{(1)\grav}
       + \frac{1}{\cseps} S_{n}^{(2)\grav} \Bigr) M_{n-1}^\tree 
 + \Ord(\cseps^0)\,,
\label{GeneralGravSoftLimit}
\end{equation}
where $M_{n}^\tree$ is the $n$-point amplitude and $M_{n-1}^\tree$ is
the $(n-1)$-point amplitude obtained by removing the soft leg $n$.
The connection of the two scalings is through little-group scaling.
The proof of universality~\cite{CachazoStrominger} of the 
subleading soft behavior~(\ref{SubleadingSoftGrav}) relies on all
contributions arising from factorizations on $1/(k_a + k_n)^2$
propagators in the soft kinematics~(\ref{SoftKinematics}), as
illustrated in \fig{TreeSoftFigure}.

Some care is needed to interpret the soft behavior in
\eqn{GeneralGravSoftLimit} because the $n$-point kinematics of the
amplitude on the left-hand side of the equation is not the same as the
$(n-1)$-point kinematics normally used to define the amplitude on the
right-hand side of the equation. This becomes an issue for the
subleading soft terms because of feed down from leading terms to
subleading ones, depending on the precise prescription.
The prescription chosen by Cachazo and Strominger is
to explicitly impose $n$-point momentum conservation on the amplitude
on the left-hand side and $(n-1)$-point momentum conservation on the
amplitude on the right-hand side.  This constraint is conveniently
implemented via
\begin{equation}
\tlambda_1 = - \sum_{i = 3}^m \frac{\spa2.i}{\spa2.1} \tlambda_i\,, \hskip 3 cm 
\tlambda_2 = - \sum_{i = 3}^m \frac{\spa1.i}{\spa1.2} \tlambda_i\,,
\end{equation} 
so that $\sum_{i = 1}^m \lambda_i \tlambda_i = 0$. 
This constraint is imposed on the amplitudes on the left-hand
side of \eqn{GeneralGravSoftLimit} with $m=n$ and on the right-hand
side with $m=n-1$. 

For our loop-level study, we use a different prescription.  We
interpret the expressions on both sides of \eqn{SoftKinematicsReal} as
carrying the {\it same} $n$-point kinematics, without needing to apply
any additional constraints on the kinematics. The advantage is that
this prevents complicated terms from feeding down from higher- to
lower-order terms in the soft expansion, which would obscure the
structure at loop level.  This change in prescription
effectively shifts contributions between different orders in the
expansion.\footnote{We
  numerically confirmed in many examples that the two prescriptions
  give identical results through $\Ord(\cseps)$ in
  \eqn{GeneralGravSoftLimitReal}.}

\subsection{Soft gluons}

Following the same derivation as for gravitons, tree-level Yang-Mills
amplitudes also have a universal subleading soft
behavior~\cite{SoftGluonProof}. If we scale $\lambda_n \rightarrow
\cseps\lambda_n$, the color-ordered amplitude behaves as
\begin{equation}
A_{n}^\tree \rightarrow
\Bigl(  \frac{1}{\cseps^2} S^{(0)}_{n\,\glue} 
      + \frac{1}{\cseps} S^{(1)}_{n\,\glue} \Bigr) A_{n-1}^\tree\,,
\label{YMSoftTree}
\end{equation}
where the leading soft factor is
\begin{equation}
S_{n\,\glue} ^{(0)} = 
 -\frac{k_{n-1}\cdot\pol_n}{\sqrt{2}\,k_{n-1} \cdot k_n}
+ \frac{k_{1}\cdot\pol_n}{\sqrt{2}\,k_1 \cdot k_n} \,.
\end{equation}
The subleading one is
\begin{equation}
S_{n\,\glue} ^{(1)}=
\frac{k_{n\mu} \pol_{n\nu} J_{n-1}^{\mu\nu}}{\sqrt{2}\,k_{n-1}\cdot k_n}
- \frac{k_{n\mu} \pol_{n\nu} J_1^{\mu\nu} }{\sqrt{2}\,k_1\cdot k_n}\,.
\label{MsoftSubleading}
\end{equation}
Again we have suppressed the coupling constants.
Using spinor-helicity, the plus-helicity gluon leading soft
 factor is
\begin{equation}
S_{n\,\glue}^{(0)}=\frac{\langle (n-1)\,1\rangle}
               {\langle (n-1)\,n\rangle\langle n\,1\rangle}\,,
\end{equation}
while the subleading operator is 
\begin{align}
S_{n\,\glue} ^{(1)}=\frac{1}{\langle(n-1)\,n\rangle}\tilde{\lambda}_n^{\dot{\alpha}}
\frac{\partial}{\partial\tilde{\lambda}_{n-1}^{\dot{\alpha}}}
- \frac{1}{\langle 1\,n\rangle}\tilde{\lambda}_n^{\dot{\alpha}}
  \frac{\partial}{\partial{\tilde{\lambda}}_1^{\dot{\alpha}}}\,.
\end{align}
An earlier description was given in Ref.~\cite{WhiteYM}.

%%%%%%%%%%%%%%%%%%

\section{One-loop corrections to subleading soft behavior}
\label{LoopSection}
 
As shown by Weinberg~\cite{Weinberg}, the leading soft-graviton
behavior has no higher-loop corrections.  In
Ref.~\cite{CachazoStrominger}, Cachazo and Strominger demonstrated
that their proposed theorem for subleading soft-graviton behavior
holds at tree level. 

Here, we demonstrate that there are nontrivial loop corrections for
the subleading soft-graviton behavior analogous to the ones that
appear in QCD for the leading soft terms, using the standard
definition of soft limits in dimensional regularization.  As in QCD, loop
corrections linked to infrared divergences necessarily appear because
of mismatches in the logarithms of the infrared singularities at $n$
and $n-1$ points.  Divergences require a regulator which can break
symmetries at the quantum level.  In this sense, we can think of the
loop corrections as due to an anomaly in the underlying symmetry.  Its
origin is similar to the twistor-space holomorphic
anomaly~\cite{HolomorphicAnomaly}, where extra contributions arise in
regions of loop integration that are singular.

In general, the structure of the loop corrections to soft behavior is
entangled with the infrared divergences.  This phenomenon is familiar
in QCD~\cite{BernChalmers,OneLoopSoftBern}, so we discuss this case
first before turning to gravity.  Besides corrections that arise from
infrared singularities, we will find that there are other loop
corrections due to nontrivial factorization
properties~\cite{SingleMinus, NontrivialFactorization,
  DunbarFactorization}, even for infrared-finite one-loop amplitudes.

\subsection{One-loop corrections to soft-gluon behavior}
\label{OneloopGluonSubsection}

%%%%%%%%% FIGURE %%%%%%%%%%%%%%%                                              
\begin{figure}[tb]
\begin{center}
\vskip .7 cm 
\includegraphics[scale=.4]{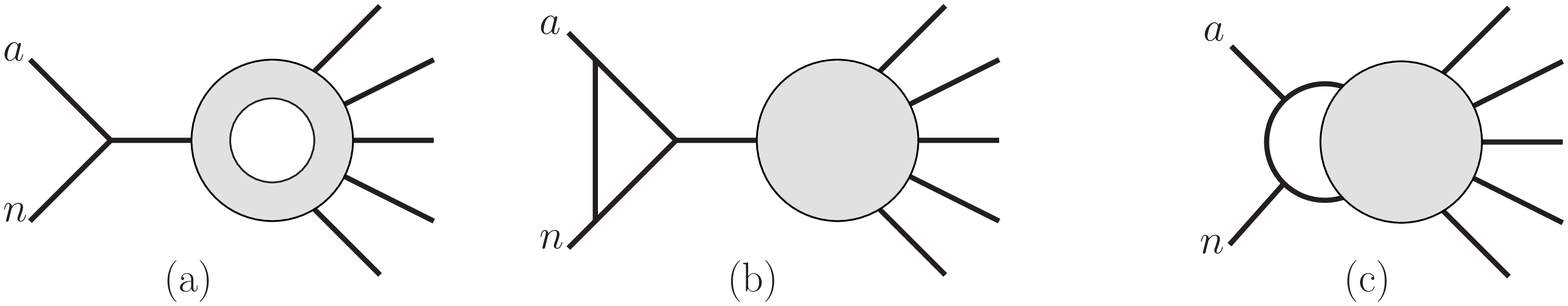}
\end{center}
\vskip -.7 cm 
\caption[a]{\small At one loop, the simple tree-level soft behavior (a)
  is corrected by factorizing (b) and nonfactorizing (c)
  contributions~\cite{OneLoopSoftBern}.  In gravity, the corrections
  are suppressed by factors of the soft momentum $k_n$, but they affect
  the subleading behavior.
\label{LoopSoftFigure}
}
\end{figure}
%%%%%%%%%%%%%%%%%%%%%%%%%%%%%%%%

In general, loop-level factorization properties of gauge theories are
surprisingly nontrivial, in part, because of their entanglement with
infrared singularities~\cite{BernChalmers}.  This causes naive notions
of factorization in soft and other kinematic limits to break down; in
massless gauge theories, one can obtain kinematic poles also from the
loop integration. However, because the infrared singularities have a
universal behavior, they offer a simple means for studying soft limits
of loop amplitudes with an arbitrary number of external legs.

\Fig{LoopSoftFigure} shows the types of contributions to the one-loop
soft behavior when the amplitude is represented in terms of the
standard covariant basis of integrals.  These consist of
``factorizing'' contributions, illustrated in \fig{LoopSoftFigure}(b),
and ``nonfactorizing'' contributions, illustrated in
\fig{LoopSoftFigure}(c).\footnote{In light-cone gauge or the unitarity
  approach, by introducing light-cone denominators containing a
  reference momentum, one can push all contributions into factorizing
  diagrams~\cite{StermanSoft,OneLoopSoftKosower}.}  The nonfactorizing
contributions arise from poles in the $S$-matrix coming from loop
integration and not directly from propagators, as illustrated in
\fig{LoopSoftFigure}(c).

\begin{figure}[tb]
\begin{center}
\vskip .7 cm 
\includegraphics[scale=.4]{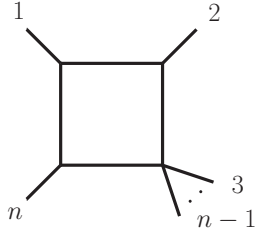}
\end{center}
\vskip -.7 cm 
\caption[a]{\small An example of an integral that has a ``nonfactorizing''
kinematic pole that contributes to the soft behavior.
\label{Box1mFigure}
}
\end{figure}

As a simple example, consider the single-external-mass box integral,
displayed in \fig{Box1mFigure}.  This is one of the basis integrals 
for  one-loop amplitudes. The infrared-divergent terms of this
integral are~\cite{BasisIntegrals} 
\begin{equation}
I_4^{\rm 1m} = 
 \frac{2i\,\cg}{s_{n 1} s_{12}} \, \biggl[ \frac{1}{\eps^2} \biggl(
\Bigl(\frac{\mu^2}{-s_{n1}}\Bigr)^{\eps} 
  + \Bigl(\frac{\mu^2} {-s_{12}} \Bigr)^{\eps}
  - \Bigl(\frac{\mu^2}{-s_{n12}}\Bigr)^{\eps}  \biggr)
+ {\rm finite} \Bigr] \,,
\end{equation}
where the labels correspond to those in \fig{Box1mFigure}.  We also have
\begin{align}
\cg=\frac{1}{(4\pi)^{2-\epsilon}}\frac{\Gamma(1+\epsilon)\Gamma^2(1-\epsilon)}{\Gamma(1-2\epsilon)}\,,\hspace{1cm}s_{i_1i_2\cdots i_j}=(k_{i_1}+k_{i_2}+\cdots +k_{i_j})^2\,.
\label{cgdef}
\end{align}
When leg $n$ goes soft, the integral has a $1/s_{n 1}$ kinematic pole
from the prefactor.  While one might expect such poles to cancel out of
amplitudes, they, in fact, remain due to their
entanglement with infrared singularities.  However, this link ensures
that they have a regular pattern.  In general, these nonfactorizing
contributions need to be accounted for in loop-level soft behavior
and other factorization limits in gauge theories. The same holds for
the subleading soft behavior of gravity amplitudes.

A one-loop $n$-gluon amplitude in QCD
has  ultraviolet and infrared singularities 
given by~\cite{OneLoopDivergence,BernChalmers}
\begin{equation}
A_n^\oneloop(1,2, \cdots, n)\Bigr|_\div =  - \frac{1}{\eps^2}
A_n^\tree(1,2, \cdots, n) \sigma^\glue_n \,,
\label{SingularIR}
\end{equation}
where 
\begin{equation}
\sigma^\glue_n =   \cg\biggl[ \sum_{j=1}^{n} 
  \biggl( \frac{\mu^{2}}{-s_{j,j+1}} \biggr)^\eps
  + {2}{\eps} \biggl(\frac{11}{6} - \frac{1}{3} \frac{n_{\!f}}{N_c}
                  - \frac{1}{6} \frac{n_s}{N_c} \biggr)  \biggr]\,.
    \label{eq:sigmaYM}
\end{equation}
In this expression, $n_f$ is the number of quark flavors, $n_s$ is the
number of scalar flavors (zero in QCD) and $N_{c}$ is the number of
colors.  Here, $\eps = (4-D)/2$ is the dimensional-regularization
parameter, and $\mu^{2}$ is the usual dimensional-regularization
scale.  It turns out that it is best to work with unrenormalized
amplitudes containing also ultraviolet divergences because the
mismatch in the number of coupling constants at $n$ and $n-1$ points
causes an additional (trivial) discontinuity in the soft behavior. By
working with unrenormalized amplitudes, we avoid this.  A key property
of Eq.~\eqref{eq:sigmaYM} is that the terms depending on the number of
quark and scalar flavors is independent of the number of external
gluons.  The terms in the summation arise from soft-gluon
singularities in the loop integration.  In general, the expression in
\eqn{eq:sigmaYM} should be interpreted as being series expanded in
$\eps$, since terms beyond $\Ord(\eps^0)$ that are usually not
computed can mix nontrivially with these.

%%%%%%%%% FIGURE %%%%%%%%%%%%%%%                                              
%%%%%%%%%%%%%%%%%%%%%%%%%%%%%%%%

Consider the soft limit of the singular parts of the gauge-theory
amplitude (\ref{SingularIR}).  The tree prefactor obeys the simple
soft behavior given in \eqn{YMSoftTree}.  The infrared singularities, however, 
have a mismatch between $n$ points and $n-1$ points:
\begin{equation}
\sigma^\glue_n = \sigma^\glue _{n-1}
   +  \sigma_n^{\prime \glue} + \Ord(\eps^{2})\,,
\end{equation}
where
\begin{equation}
\sigma_n^{\prime\glue} = \cg \biggl(1 +
   \eps \log\biggl(\frac{-\mu^2 s_{(n-1) 1}}{s_{(n-1) n} s_{n 1}}
          \biggr) \biggr) \,.
\end{equation}
It turns out that this mismatch can be used to deduce
the complete one-loop corrections to the
leading soft factor by matching the infrared
discontinuities in the basis integrals to the infrared discontinuities
in the amplitude~\cite{OneLoopSoftBern}. 

The leading soft behavior of an $n$-gluon amplitude with any matter
content for $\lambda_n \rightarrow \cseps \lambda_{n} $ is
then~\cite{OneLoopSoftBern,OneLoopSoftKosower}
\begin{equation}
A_n^\oneloop \rightarrow S_{n\,\glue}^{(0)} A_{n-1}^\oneloop+  S_{n\,\glue}^{(0) \oneloop} A_{n-1}^\tree\,,
\end{equation}
where the leading one-loop soft correction function is
\begin{eqnarray}
S_{n\,\glue}^{(0)\oneloop} &=& -S_{n\,\glue}^{(0)} \frac{\cg}{\eps^2} 
\bigg(\frac{-\mu^2 s_{(n-1) 1}}{s_{(n-1)n} s_{n1}} \biggr)^{\eps}
               \frac{\pi \eps}{\sin(\pi \eps)}\nonumber \\
&=& -S_{n\,\glue}^{(0)} \cg  \biggl( \frac{1}{\eps^2} 
   + \frac{1}{\eps} \log\biggl(\frac{-\mu^2 s_{(n-1) 1}}{\cseps^2s_{(n-1) n} s_{n 1}}
            \biggr)  
  + \frac{1}{2} 
   \log^2\biggl(\frac{-\mu^2 s_{(n-1) 1}}{\cseps^2 s_{(n-1) n} s_{n 1}}\biggr)
        + \frac{\pi^2}{6} \biggr) \nonumber\\
&& \null \hskip 2 cm 
  + \Ord(\eps)\,.
\label{OneLoopLeadingSoftFunction}
\end{eqnarray}
The form on the first line is valid to all orders in $\eps$.  In
applying this equation, it is important to first expand in $\eps$
prior taking the soft limit.

Now consider the subleading soft terms. 
Taking the divergent part of the one-loop amplitude to have
a soft limit of the form, 
\begin{equation}
A_{n}^\oneloop \Bigr|_{\div} \rightarrow
\Bigl( \frac{1}{\cseps^2} S^{(0)}_{n\,\glue}
      + \frac{1}{\cseps} S^{(1)}_{n\,\glue} \Bigr) A_{n-1}^\oneloop \Bigr|_\div +
 \Bigl( \frac{1}{\cseps^2} S^{(0)\oneloop}_{n\,\glue}
      + \frac{1}{\cseps} S^{(1)\oneloop}_{n\,\glue} \Bigr) A_{n-1}^\tree\Bigr|_\div \,,
\label{YMOneloopSoft}
\end{equation}
we then solve for the divergent parts of the one-loop corrections to
the soft operators, denoted by $S^{(i)\,\oneloop}_{n\,\glue}$.  We do
so by comparing the soft expansion of the left-hand side of
\eqn{YMOneloopSoft} to the terms on the right-hand side.  Applying
$S^{(1)}_{n\,\glue}$ to the infrared singularity of the $(n-1)$-point
amplitude gives
\begin{eqnarray}
S^{(1)}_{n\,\glue} \sigma^\glue_{n-1} &=& 
-\cg \eps 
\Bigl(\frac{\spb1.n}{\spb1.{(n-1)} \spa{(n-1)}.n} -  \frac{\spb{(n-1)}.n} {\spb{(n-1)}.1\spa{1}.n }
   \nonumber \\
&& \null \hskip 3 cm 
 + \frac{\spb{(n-2)}.n} {\spb{(n-2)}.{(n-1)}\spa{(n-1)}.n }-  \frac{\spb{2}.n} {\spb{2}.{1} \spa{1}.n} \Bigr)\,,
\label{S1YMResult}
\end{eqnarray}
where we use the form of $\sigma^\glue_{n-1}$ exactly as it appears in
\eqn{eq:sigmaYM} without any additional momentum-conservation
relations imposed.  Taking the one-loop correction to the subleading soft
function to be
\begin{eqnarray}
S^{(1)\oneloop}_{n\,\glue} & = &
-  \frac{1}{\eps^2} \biggl[
    \sigma_n^{\prime \glue} S^{(1)}_{n\,\glue}
   - \Big(S^{(1)}_{n\,\glue} \sigma^\glue_{n-1}\Bigr)
 \biggr]
  + \Ord(\eps^0) \,,
\end{eqnarray}
we find that \eqn{YMOneloopSoft} holds.  The simple form of the
correction relies on using the specific form for $S^{(1)}\sigma_{n-1}$ in 
\eqn{S1YMResult}.  We also interpret both sides of \eqn{YMOneloopSoft} as
having the same $n$-point kinematics.

It would be important to understand the infrared-finite terms as well.
These also have nontrivial corrections.  For the case of the
infrared-finite identical-helicity one-loop amplitudes~\cite{AllPlus},
numerical analysis through 30 points shows that the amplitudes behave
exactly as tree-level amplitudes with no nontrivial corrections.
However, the one-loop amplitudes with a single minus
helicity~\cite{SingleMinus} have nontrivial subleading soft behavior.
As an example, consider the one-loop five-gluon
amplitude~\cite{fivePointOneLoop,SingleMinus},
\begin{align}
A_5^\oneloop(1^-,2^+,3^+,4^+,5^+)=\frac{i}{48\pi^2}
\frac{1}{\spa3.4^2} \biggl[-\frac{\spb2.5^3}{\spb1.2 \spb5.1} 
  + \frac{\spa1.4^3\spb4.5 \spa3.5}{\spa1.2 \spa2.3 \spa4.5^2} 
 - \frac{\spa1.3^3\spb3.2\spa4.2}{\spa1.5\spa5.4\spa3.2^2} \biggr] \,,
\label{SingleMinusYM}
\end{align}
as the momentum of leg 5 becomes soft.  The four-point
one-loop single-minus-helicity amplitude is~\cite{FourPointSingleMinus}
\begin{align}
A_4^\oneloop(1^-,2^+,3^+,4^+)=\frac{i}{48\pi^2}\frac{\spa2.4 \spb2.4^3}
                  {\spb1.2 \spa2.3 \spa3.4 \spb4.1}\,.
\label{SingleMinusYMFourPt}
\end{align}
Applying the tree-level operators to the four-point amplitude, 
as in Eq.~\eqref{YMSoftTree}, yields
\begin{align}
&\Bigl(  \frac{1}{\cseps^2} S^{(0)}_{n\,\glue} 
      + \frac{1}{\cseps} S^{(1)}_{n\,\glue} \Bigr)
 A_{4}^\oneloop(1^-,2^+,3^+,4^+) \notag \\
   &\hspace{4cm}=\frac{i}{48\pi^2}\frac{\spa1.3^3 \spa2.4 \spb1.2}
  {\spa2.3^2\spa3.4^3} \left(\frac{1}{\delta^2}
    \frac{\spa4.1}{\spa4.5 \spa5.1} 
   +\frac{1}{\delta} \frac{\spb5.2}{\spa5.1 \spb1.2} \right)\,.
\label{FourPTOperator}
\end{align}
After applying the operators, we applied five-point momentum
conservation to remove the anti-holomorphic spinors
$\tilde{\lambda}_3$, $\tilde{\lambda}_4$.\footnote{We note that the
  momentum-conservation prescription of Ref.~\cite{CachazoStrominger}
  gives the same conclusion.}  This facilitates comparison with the
soft limit of the five-point amplitude~\eqref{SingleMinusYM}. With the
same constraints applied, this is given by
\begin{align}
A_5^\oneloop(1^-,2^+,3^+,4^+,5^+)&\rightarrow
\frac{i}{48\pi^2}\bigg[\frac{\spa1.3^3 \spa2.4 \spb1.2}{\spa2.3^2 \spa3.4^3}
\left(\frac{1}{\delta^2} \frac{\spa4.1}{\spa4.5 \spa5.1} 
 + \frac{1}{\delta} \frac{\spb5.2}{\spa5.1 \spb1.2}\right) \notag \\
&\hspace{2cm}\null 
+\frac{1}{\delta}\frac{\spa1.4^3\spa3.5}
  {\spa1.2 \spa2.3 \spa 3.4^3 \spa4.5^2}
  (\spa1.3 \spb1.5 + \spa2.3 \spb2.5)\bigg]\,.
\label{FivePointSoft}
\end{align}
While the leading order pieces are identical, the subleading pieces differ
in \eqns{FourPTOperator}{FivePointSoft}.

The nontrivial behavior of the single-minus-helicity amplitudes is not
surprising given that they contain nontrivial complex poles that
cannot be interpreted as a straightforward factorization.  In general,
nonsupersymmetric gauge-theory loop amplitudes contain such nontrivial
poles.  This phenomenon complicates the construction of gauge and
gravity loop amplitudes from their poles and has been described in
some detail in
Refs.~\cite{NontrivialFactorization,DunbarFactorization}.  We leave
the discussion of such infrared-finite contributions to the future.

\subsection{One-loop corrections to soft-graviton behavior}
\label{OneloopGravitonSubsection}

Applying a similar analysis, it is straightforward to see that one-loop
corrections to the subleading soft-graviton behavior do not vanish
because of mismatched logarithms in the infrared singularities.
At one loop, the $n$-graviton amplitude contains the
dimensionally-regularized infrared-singular
terms~\cite{DunbarNorridge,GravityIR},
\begin{equation}
M_n^\oneloop\Bigr|_\div = \frac{\sigma_n^\grav}{\epsilon} M_n^\tree\,,
\label{OneLoopIRGrav}
\end{equation}
where $M_n^\tree$ is the $n$-graviton tree amplitude, and 
\begin{equation}
\sigma_n= - \cg
\sum_{i=1}^{n-1}\sum_{j=i+1}^ns_{ij} \log\Bigl(\frac{\mu^2}{-s_{ij}}\Bigr)\,,
\label{SigmaGrav}
\end{equation}
where $\cg$ is defined in \eqn{cgdef}.
As in QCD, the logarithms that appear at $n$ points are not identical 
to the ones appearing at $(n-1)$ points. The logarithms in 
the infrared singularity that 
differ between an $n$- and $(n-1)$-graviton amplitude are 
\begin{equation}
\sigma'_{n} = - \cg
 \sum_{i=1}^{n-1} s_{in}
\log\Bigl(\frac{\mu^2}{-s_{in}}\Bigr) \,.
\label{SigmaPrime}
\end{equation}
While this mismatch does not affect the leading soft behavior because of  
the suppression from the $s_{in}$ factors, it does affect
subleading terms. 

By absorbing the mismatches into corrections to the subleading soft
operator, we find that in the soft limit $\lambda_n \rightarrow \cseps
\lambda_n$, the infrared singular terms behave as
\begin{equation}
M_n^\oneloop \Bigr|_\div  \rightarrow  \biggl(
\frac{S_{n\,\grav}^{(0)}}{\cseps^3} + \frac{S_{n\,\grav}^{(1)}}{\cseps^2}
+ \frac{S_{n\,\grav}^{(2)}}{\cseps}\biggr) M_{n-1}^\oneloop \Bigr|_\div
+\biggl(\frac{S_{n\,\grav}^{(1)\,\oneloop}}{\cseps^2} +
       \frac{S_{n\,\grav}^{(2)\,\oneloop}}{\cseps}  \biggr)
       M_{n-1}^\tree \Bigr|_\div \,,
 \label{OneloopBehavior}
\end{equation}
where 
\begin{align}
S_{n\,\grav}^{(0)\,\oneloop} \Bigr|_\div & = 0
\,,\nonumber \\
S_{n\,\grav}^{(1)\,\oneloop} \Bigr|_\div & = \frac{1}{\eps} \biggl[
   \sigma_n' S_{n\,\grav}^{(0)}
   - \Bigl(S_{n\,\grav}^{(1)} \sigma^\grav_{n-1} \Bigr) \biggr] 
\,,\nonumber \\
S_{n\,\grav}^{(2)\,\oneloop}\Bigr|_\div &= \frac{1}{\eps} \biggl[
  \sigma_n' S_{n\,\grav}^{(1)} 
 - \Bigl(S_{n\,\grav}^{(2)} \sigma^\grav_{n-1} \Bigr)
    + \sum_{i=1}^{n-1} \frac{\spb{n}.{i}}{\spa{n}.{i}} \,
 \biggl(  \tlambda^{\dot\alpha}_n  \frac{\partial\sigma_{n-1}^\grav} 
    {\partial \tlambda^{\dot\alpha}_i}   \biggr)\,{\tlambda^{\dot\beta}_n}
          \frac{\partial} {\partial \tlambda^{\dot\beta}_i }
\biggr]  \,.
\label{OneLoopSoftFunctions}
 \end{align}
Similar to the gauge-theory case, the simple form of these corrections to
the subleading soft operators relies on using the form of $\sigma_{n-1}$
obtained from \eqn{SigmaGrav} with no additional momentum-conservation
relations imposed.  We again also interpret both sides of
\eqn{OneloopBehavior} as having the same $n$-point kinematics.  As in
QCD, it is important to follow the standard procedure of first series
expanding the amplitude in $\eps$ prior to taking soft limits.

We have checked numerically through 10 points that the infrared-finite
identical-helicity graviton amplitudes~\cite{AllPlusGrav} satisfy the
same subleading soft behavior as the tree amplitudes.  However, more
generally we expect a more complicated behavior due to the nontrivial
factorization properties of loop amplitudes~\cite{SingleMinus,
  NontrivialFactorization}. Such nontrivial factorization properties
have been discussed for gravity theories in
Refs.~\cite{DunbarSingleMinusFive,DunbarFactorization}.  Indeed, by
numerically analyzing the infrared-finite one-loop five-graviton
amplitude with a single minus helicity from
Ref.~\cite{DunbarSingleMinusFive} and the one-loop four-graviton
amplitude with a single minus helicity from
Ref.~\cite{BernDunbarShimada}, we find that the second subleading soft
behavior has nontrivial corrections.  We leave a discussion of the
infrared-finite corrections to the graviton soft behavior to the future.

\section{All loop order behavior of soft gravitons}
\label{AllLoopSection}

As we demonstrated in the previous section, the subleading soft
behavior has loop corrections.  In this section, we argue that the
first subleading soft behavior has no corrections beyond one loop and
that the second subleading behavior has no corrections beyond two
loops.

%%%%%%%%% FIGURE %%%%%%%%%%%%%%%                                              
\begin{figure}[tb]
\begin{center}
\vskip .7 cm 
\includegraphics[scale=.4]{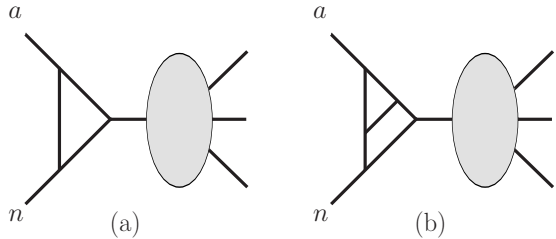}
\end{center}
\vskip -.7 cm 
\caption[a]{\small Sample factorizing (a) one- and (b) two-loop 
contributions to the soft behavior.
\label{FactorizingFigure}
}
\end{figure}
%%%%%%%%%%%%%%%%%%%%%%%%%%%%%%%%

%%%%%%%%% FIGURE %%%%%%%%%%%%%%%                                              
\begin{figure}[tb]
\begin{center}
\vskip .7 cm 
\includegraphics[scale=.55]{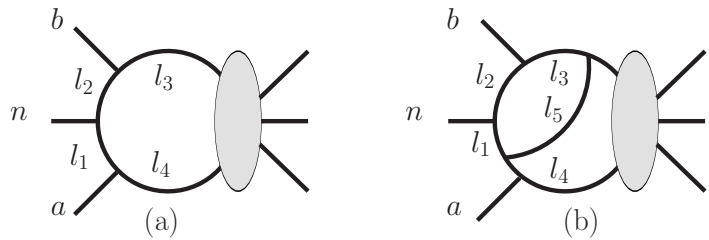}
\end{center}
\vskip -.7 cm 
\caption[a]{\small Sample nonfactorizing (a) one- and (b) two-loop 
contributions to the soft behavior.
\label{NonFactorizingFigure}
}
\end{figure}
%%%%%%%%%%%%%%%%%%%%%%%%%%%%%%%%

\subsection{General considerations}

The all-loop leading soft-graviton behavior has been
discussed in some detail in Section~{5.2} of
Ref.~\cite{OneLoopMHVGrav}.  Here we follow this discussion for the
subleading behavior. As already noted for gauge theory, potential
contributions to the soft behavior can be divided into ``factorizing''
and ``nonfactorizing'' contributions~\cite{BernChalmers} when the
amplitude is expressed in terms of covariant Feynman integrals. We
consider these types of contributions in turn.

The factorizing contributions of the type displayed in
\fig{FactorizingFigure} depend on the soft momentum $k_n$ and one
additional momentum $k_a$. After the Lorentz indices of polarization
tensors are contracted, no other Lorentz invariants are present other
than $s_{na}$.  By dimensional analysis, the $L$-loop correction
contains an additional factor $\kappa^{2L}$ of the gravitational
coupling relative to the tree-level contribution in
\fig{TreeSoftFigure}, and therefore must contain relative factors of
$s_{na}^L$.  This gives a suppression of one soft momentum $k_n$ for
each additional loop.

The nonfactorizing contributions displayed in
\fig{NonFactorizingFigure} have a similar suppression. The
nonfactorizing contributions arise in regions where loop momenta
become soft in addition to the external soft leg.  For example, in the
one-loop case displayed in \fig{NonFactorizingFigure}(a), as $k_n
\rightarrow 0$, we must also have the loop momentum go as $l_1
\rightarrow 0$ in order to obtain a nonfactorizing contribution to the
soft behavior; otherwise, there would be no large contribution for
$k_n \rightarrow 0$, or equivalently for $\lambda_n \rightarrow 0$.
In this region, $l_2 = l_1 - k_n$, $l_3 = l_1 - k_n - k_b$ and $l_4 =
l_1 + k_a$ also all become small.  After integration, this leads to
potential kinematic poles in $s_{an}$ or $s_{bn}$, or equivalently in
$\lambda_n$.  However, because gravity has an extra power of soft
momentum, either $k_n$ or $l_1$ in the vertex attaching leg $n$ to the
loop will suppress the pole.  Similarly, at two loops, illustrated in
\fig{NonFactorizingFigure}(b), potential contributions arise when
additional loop momenta become soft, in this case $l_5$.  Once again,
the dimensionful coupling ensures that there will be additional
factors of soft momenta in the numerator.  More generally, after
integration, we get an additional $L$ factors of $s_{jn}$ compared to
the gauge-theory case, where $j$ can be any momentum in the amplitude.

The net affect effect is that there are no loop corrections to the
leading soft behavior, no corrections beyond one loop for the first
subleading soft behavior, and no corrections beyond two loops for the
second subleading soft behavior.  We therefore expect the general form
of the $L$-loop behavior for a plus-helicity graviton with
$\lambda_n \rightarrow \cseps\lambda_n$ to have no loop corrections
beyond two loops.

\subsection{All loop behavior of leading infrared singularities}

Since there should be no corrections beyond two loops, we expect that
the $L$-loop leading infrared-divergent terms should behave in the
soft limit as
\begin{eqnarray}
M_n^\Lloop\Bigr|_{\ldiv} & \hskip -.2 cm\rightarrow & \biggl(
\frac{S_{n\,\grav}^{(0)}}{\cseps^3} + \frac{S_{n\,\grav}^{(1)}}{\cseps^2} 
+ \frac{S_{n\,\grav}^{(2)}}{\cseps}\biggr) M_{n-1}^\Lloop \Bigr|_{\ldiv}  \nonumber\\
&& \null \hskip 2 cm
+\biggl(\frac{S_{n\,\grav}^{(1)\,\oneloop}}{\cseps^2} +
       \frac{S_{n\,\grav}^{(2)\,\oneloop}}{\cseps}  \biggr)
       M_{n-1}^\Lmoneloop \Bigr|_{\ldiv} \nonumber \\
&& \null \hskip 2 cm
+ \frac{S_{n\,\grav}^{(2)\,\twoloop}}{\cseps}
       M_{n-1}^\Lmtwoloop \Bigr|_{\ldiv} \,.
\label{LloopBehavior}
\end{eqnarray}
We check this using the known all-loop-order form of infrared
singularities in gravity theories~\cite{Weinberg,GravityIR}. The
infrared singularities of gravity amplitudes are given by
\begin{equation}
M_n = {\cal S}_n {\cal H}_n\,,
\end{equation}
where $M_n$ is a gravity amplitude valid to all loop orders and ${\cal
  H}_n$ is the infrared-finite hard function. The all-loop 
infrared singularity function is a simple exponentiation of the
one-loop function~(\ref{OneLoopIRGrav}):
\begin{equation}
{\cal S}_n = \exp\Big( \frac{\sigma_n}{\eps} \Bigr)\,.
\end{equation}
From this equation, we see that the 
leading infrared singularity at $L$ loops is simply given in terms
of the tree amplitude:
\begin{equation}
M_n^\Lloop \Bigr|_{\ldiv} = \frac{1}{L!} 
    \left(\frac{\sigma_n}{\epsilon}\right)^L M_n^\tree\,.
\end{equation}

This gives us a simple means for testing \eqn{LloopBehavior} and also 
for finding the leading infrared-singular part of the two-loop operator,
$S_{n\,\grav}^{(2)\,\twoloop}$.  We do so 
by taking the difference of the soft expansion on both sides
of \eqn{LloopBehavior} and using the previously determined operators
in \eqn{OneLoopSoftFunctions}.
We need the soft expansion of the leading infrared-singular part of $M_n^\Lloop$, given by 
\begin{align}
\label{expansion}
\frac{\sigma_{n}^{L}}{L!}M_{n}^{\tree}
&\rightarrow
	\frac{\left(\sigma_{n-1}+\cseps\sigma_{n}^{\prime}\right)^{L}}{L!}
  \biggl(\frac{S_{n\,\grav}^{(0)}}{\cseps^3} + \frac{S_{n\,\grav}^{(1)}}{\cseps^2}
	+ \frac{S_{n\,\grav}^{(2)}}{\cseps}\biggr) M_{n-1}^\tree\,,
\end{align}
where $\sigma_n'$ is defined in \eqn{SigmaPrime}.  We also need
the results of acting on $(\sigma_{n-1}^{L}/L!)M_{n-1}^{\tree}$ with the tree-level
soft operators, 
\begin{align}
\label{operator}
	&\biggl(\frac{S_{n\,\grav}^{(0)}}{\cseps^3} 
		+ \frac{S_{n\,\grav}^{(1)}}{\cseps^2}
		+ \frac{S_{n\,\grav}^{(2)}}{\cseps}\biggr) 
	\frac{\sigma^{L}_{n-1}}{L!}M_{n-1}^{\tree} \,.
\end{align}
Evaluating these, we deduce the leading
infrared-divergent contribution to the two-loop soft operator to be
\begin{equation}
S_{n\,\grav}^{(2)\,\twoloop} \Bigr|_\ldiv = 
	\frac{1}{\eps^{2}}\Biggl[
     \frac{1}{2} \left(\sigma_{n}^{\prime}\right)^{2}S_{n\,\grav}^{(0)}
     - \sigma_n' \Bigl( S_{n\,\grav}^{(1)} \sigma_{n-1} \Bigr)
	-\left(\frac{1}{2}\sum_{i=1}^{n-1}\frac{\spb{n}.i}{\spa{n}.i}\,
	\left(\tilde{\lambda}^{\dot{\alpha}}_{n}\frac{\partial\sigma_{n-1}}{\partial\tilde{\lambda}^{\dot{\alpha}}_{i}}\right)^{2}\right)
	\Biggr] \,.
\end{equation}
The lack of higher-loop corrections to the soft operators is a
consequence of the fact that they are suppressed by additional powers
of the soft momentum. As before, the form of $\sigma_{n-1}$ in the
correction must be specifically as given in \eqn{SigmaGrav}.

\section{Conclusions}
\label{ConclusionSection}

Recently a generalization of Weinberg's soft-graviton theorem for the
subleading behavior was
proposed~\cite{StromingerNotes,CachazoStrominger}.  (See also previous
work from White~\cite{WhiteGrav}.) Here we showed that, unlike the
leading soft-graviton behavior, the subleading soft behavior requires
loop corrections. In QCD, loop corrections to the leading soft
functions make up for mismatches in the infrared singularities of
$n$-point and $(n-1)$-point amplitudes.  Applying this observation to
gravity, we obtained the leading infrared-singular loop contributions
to the subleading soft-graviton operators valid to all loop
orders. This proves in a simple way that there necessarily are
nonvanishing loop corrections to soft-graviton behavior.  In addition,
in the simple example of a five-graviton amplitude with 
a single minus helicity, we found additional corrections to the second
subleading behavior, not linked to infrared singularities. These come from
the nontrivial complex factorization properties of generic loop
amplitudes~\cite{BernChalmers, SingleMinus, NontrivialFactorization,
  DunbarSingleMinusFive, DunbarFactorization}.

Following the discussion for the leading soft-graviton
behavior~\cite{Weinberg,OneLoopMHVGrav}, we argued that there are no
loop corrections to the first subleading soft behavior beyond one loop
and no new corrections to the second subleading behavior beyond two loops.
This is connected to the dimensionful coupling of gravity.  In the
regions contributing to the soft limit, an extra power of the soft
momentum is obtained for each additional loop, suppressing the
contributions.  By the third loop order, there are a sufficient number
of powers of the soft momentum to suppress further corrections to
the soft operators.

We also discussed the form of subleading corrections to the 
soft behavior in gauge theory as a warm-up for the gravity case.
It is interesting to note that the subleading soft 
behavior in QCD might be useful for improved soft-gluon approximations.

An important remaining task is to determine the loop corrections to
the general subleading soft behavior of the infrared-finite terms in both
gauge and gravity theories.  While this is simple in special cases,
such as for identical-helicity amplitudes~\cite{AllPlus,AllPlusGrav},
in general, the task is complicated by the nontrivial complex
factorization properties of loop amplitudes~\cite{BernChalmers,
  SingleMinus, NontrivialFactorization, DunbarSingleMinusFive,
  DunbarFactorization}, on top of well understood feed downs from infrared
singularities.  We leave studies of the soft behavior of
infrared-finite terms in gauge and gravity amplitudes to future work.

\section*{Added Note}

In this paper we have used the standard definition of
dimensionally-regularized soft limits where one first series expands
in the dimensional-regularization parameter before taking the soft
limit. We do so because it matches the one needed for scattering
amplitudes and associated physical processes as they are normally
computed.  After the appearance of the first version of this paper, a
new paper appeared~\cite{FreddyNew} showing that in some simple
supersymmetric examples, loop corrections to the soft operators can be
removed by altering the long-standing standard definition of soft
limits. This alteration involves keeping the
dimensional-regularization parameter finite before taking the soft
limit.

The lack of loop corrections found in the examples of
Ref.~\cite{FreddyNew} is not surprising and is a simple consequence of
the lack of discontinuities~\cite{BernChalmers,OneLoopSoftBern} with
the reordered limits.  This is connected to the well-known fact that
with a finite dimensional-regularization parameter $\eps < 0$, or
equivalently $D>4$, there are no infrared singularities. One can also
view the prescription as equivalent to taking soft limits on
integrands instead of the integrated expressions because one can push
limits through the integral when they are smooth.  (One can apply soft
limits directly at the integrand level, but that is a distinct problem
from the one for integrated amplitudes.)  As an example, we
immediately see from the first line of
\eqn{OneLoopLeadingSoftFunction} that one-loop corrections to the
leading soft function in QCD vanish for $k_n \rightarrow 0$ if we
hold $\eps <0$ fixed.

However, there are a number of reasons why it is important to use the
standard dimensional-regularization procedure of series expanding in
$\eps$ prior to taking soft~\cite{OneLoopSoftBern, OneLoopSoftKosower}
or other limits.  To be useful for obtaining cross sections, soft
limits must be compatible with cancellations of infrared singularities
between real-emission and virtual contributions.  One might imagine
keeping $\eps$ finite in both contributions in an attempt to treat
them on an equal footing.  However, the use of four-dimensional
helicity states on external legs makes this problematic. Even in the
well-understood standard definition of soft limits, one must be
careful not to violate unitarity because of the incompatible treatment
of real-emission and virtual contributions. (See for example
Ref.~\cite{KunsztIR}.)  Moreover, in QCD the modified prescription
disrupts the cancellation of leading infrared singularities when $\eps
\rightarrow 0$ because it alters the real-emission sigularities
without changing corresponding virtual ones.

Even if there were a way to avoid difficulties with real-emission
contributions, keeping $\eps$ finite in virtual contributions would
lead to serious complications as well.  In general, loop amplitudes
are computed only through a fixed order in $\eps$ because the higher
order contributions are rather complicated, except in simple
supersymmetric cases, and do not carry useful physical information for
the problem at hand. (For an example of the typical forms that 
loop amplitudes take, see
Ref.~\cite{FivePtQCD}.) 

The single-minus helicity infrared-finite amplitudes are a good
example of why it is best to series expand in $\eps$. As noted in
Sections~\ref{OneloopGluonSubsection} and
\ref{OneloopGravitonSubsection}, these amplitudes have another type of
loop correction to soft behavior coming from nontrivial complex
factorization channels and not from infrared discontinuities.  (Since
the first version of our paper appeared, He, Huang and Wen thoroughly
investigated the single-minus helicity amplitudes~\cite{HeHuang},
among other topics, confirming our finding of nontrivial loop
corrections.)  In general, such amplitudes are known only for $\eps =
0$~\cite{SingleMinus,DunbarSingleMinusFive}. It would be highly
nontrivial to obtain the higher order in $\eps$ contributions for the
purpose of attempting to prevent renormalization of the soft
operators.  Furthermore, we note that loop corrections to soft
behavior are, in fact, quite useful for understanding the analytic
structure of amplitudes and their associated physical properties.
More generally, experience shows that it is overwhelmingly simpler to
absorb complications associated with dimensional regularization into
loop corrections of soft limits rather than to deal with higher order in
$\eps$ terms in amplitudes.

Consequently, while it may be tempting to change the standard
definitions of dimensional regularization and soft limits in order to
remove loop corrections to soft operators associated with infrared
singularities, we greatly prefer the standard definitions because of
their well-understood consistency, simplicity and applicability to
problems of physical and theoretical interest.

\subsection*{Acknowledgments}

We thank Guillaume Bossard, John Joseph Carrasco, Yu-tin Huang, Henrik
Johansson and Radu Roiban for helpful discussions. We also thank Paolo
Di Vecchia, Duff Neill and Chris White for pointing out the earlier
work on subleading soft limits, as well as for helpful discussions.
This work was supported by the US Department of Energy under Award
Number DE-{S}C0009937. We also gratefully acknowledge Mani Bhaumik for
his generous support.

%%%%%%%%%%%%%%%%%%%%%%%%%

\end{document}